\pdfoutput=1
\documentclass[conference]{IEEEtran}
\ifCLASSINFOpdf
\else
\fi
\usepackage{cite}
\usepackage[cmex10]{amsmath}
\usepackage{multirow}
\usepackage{array}
\usepackage[lofdepth,lotdepth]{subfig}
\usepackage{lipsum}
\usepackage[pdftex]{graphicx}
\usepackage{mathtools}
\usepackage{cuted}
\usepackage{breqn}
\usepackage{graphicx}
\usepackage{gensymb}
\usepackage[english]{babel}
\usepackage[autostyle]{csquotes}
\usepackage[acronym,toc,shortcuts]{glossaries}
\usepackage{caption}
\usepackage{comment}
\usepackage{siunitx}
\usepackage{textgreek}
\usepackage{dblfloatfix}

\makeglossaries
\newacronym{ACO-OFDM}{ACO-OFDM}{Asymmetrically Clipped Optical OFDM}
\newacronym{ADAS}{ADAS}{advanced driver-assistance systems}
\newacronym{ADC}{ADC}{analog-to-digital converters}
\newacronym{APD}{APD}{avalanche photodiode}
\newacronym{AWG}{AWG}{Arbitrary Waveform Generator}
\newacronym{BER}{BER}{bit-error-rate}
\newacronym{CAN}{CAN}{controlled area network}
\newacronym{CC}{CC}{Convolutional Code}
\newacronym{CE}{CE}{channel equalization}
\newacronym{C-V2X}{C-V2X}{Cellular Vehicular to Everything Communication}
\newacronym{CLT}{CLT}{Central Limit Theorem}
\newacronym{CIR}{CIR}{Channel Impulse Response}
\newacronym{CFR}{CFR}{Channel Frequency Response}
\newacronym{CMOS}{CMOS}{Complementary Metal Oxide Semiconductor}
\newacronym{COTS}{COTS}{commercial off-the-shelf}
\newacronym{CSI}{CSI}{Channel State Information}
\newacronym{CSK}{CSK}{Color Shift Keying}
\newacronym{DC}{DC}{direct current}
\newacronym{DCO-OFDM}{DCO-OFDM}{Direct Current-Biased Optical OFDM}
\newacronym{DPIM}{DPIM}{Digital Pulse Interval Modulation}
\newacronym{DRL}{DRL}{Day Time Running Light}
\newacronym{DSRC}{DSRC}{Dedicated Short Range Communication}
\newacronym{ECU}{ECU}{electronic control units}
\newacronym{EIRP}{EIRP}{effective isotropic radiated power}
\newacronym{EM}{EM}{electromagnetic}
\newacronym{EMH}{EMH}{electromagnetic harvesters}
\newacronym{EUB}{EUB}{effective usable bandwidth}
\newacronym{Es/N0}{Es/N0}{energy per symbol to noise power spectral density ratio}
\newacronym{FEC}{FEC}{Forward Error Correction}
\newacronym{FLP}{FLP}{fast locking pattern}
\newacronym{FFT}{FFT}{Fast Fourier Transform}
\newacronym{FoV}{FoV}{field of view}
\newacronym{FPGA}{FPGA}{Field Programmable Gate Array}
\newacronym{GPSDO}{GPSDO}{Global Positioning System disciplined oscillator}
\newacronym{IF}{IF}{intermediate frequency}
\newacronym{IDFT}{IDFT}{inverse discrete Fourier transform}
\newacronym{IFFT}{IFFT}{inverse Fast Fourier Transform}
\newacronym{IM/DD}{IM/DD}{intensity modulation / direct detection}
\newacronym{IR}{IR}{infrared}
\newacronym{ISI}{ISI}{inter-symbol interference}
\newacronym{ITS}{ITS}{Intelligent Transportation Systems}
\newacronym{IVWSN}{IVWSN}{intra-vehicular wireless sensor networks}
\newacronym{LED}{LED}{light emitting diode}
\newacronym{LIDAR}{LIDAR}{Light Detection And Ranging}
\newacronym{LNA}{LNA}{low noise amplifier}
\newacronym{LoS}{LoS}{line-of-sight}
\newacronym{LTE}{LTE}{Long Term Evolution}
\newacronym{MAC}{MAC}{medium access control}
\newacronym{MCS}{MCS}{modulation coding schemes}
\newacronym{MCU}{MCU}{microcontroller units}
\newacronym{MDS}{MDS}{minimum detectable signal}
\newacronym{MPC}{MPC}{multi-path component}
\newacronym{MIMO}{MIMO}{multiple input multiple output}
\newacronym{MIP}{MIP}{"maximum interference power"}
\newacronym{mmWave}{mmWave}{millimeter wave}
\newacronym{NLoS}{NLoS}{non-line-of-sight}
\newacronym{OOK}{OOK}{On-Off Keying}
\newacronym{OFDM}{OFDM}{orthogonal frequency division multiplexing}
\newacronym{OWC}{OWC}{Optical Wireless Communication}
\newacronym{QAM}{QAM}{Quadrature Amplitude Modulation}
\newacronym{PD}{PD}{photo-detector}
\newacronym{PDP}{PDP}{power delay profile}
\newacronym{PHR}{PHR}{physical header}
\newacronym{PHY}{PHY}{physical layer}
\newacronym{PN}{PN}{pseudorandom noise}
\newacronym{PPM}{PPM}{Pulse Position Modulation}
\newacronym{PSDU}{PSDU}{physical service data unit}
\newacronym{PWM}{PWM}{pulse width modulation}
\newacronym{RADAR}{RADAR}{Radio Detection And Ranging}
\newacronym{RF}{RF}{radio frequency}
\newacronym{RFEH}{RFEH}{Radio frequency energy harvesting}
\newacronym{RLL}{RLL}{Run-Length Limited}
\newacronym{RMS}{RMS}{root mean square}
\newacronym{RS}{RS}{Reed Solomon}
\newacronym{RU}{RU}{Receiver unit}
\newacronym{RX}{RX}{receiver}
\newacronym{TX}{TX}{transmit}
\newacronym{SSA}{SSA}{signal and spectrum analyzer}
\newacronym{SDR}{SDR}{Software-Defined Radio}
\newacronym{SER}{SER}{Symbol Error Rate}
\newacronym{SHR}{SHR}{synchronization header}
\newacronym{SINR}{SINR}{signal-to-interference-and-noise ratio}
\newacronym{SNR}{SNR}{signal-to-noise ratio}
\newacronym{TDP}{TDP}{topology dependent pattern}
\newacronym{TEG}{TEG}{Thermoelectric generators}
\newacronym{TU}{TU}{transmitter unit}
\newacronym{U-OFDM}{U-OFDM}{Unipolar OFDM}
\newacronym{UWB}{UWB}{ultra wideband}
\newacronym{USRP}{USRP}{Universal Software Radio Peripheral}
\newacronym{VLC}{VLC}{visible light communication}
\newacronym{Vpp}{Vpp}{peak-to-peak voltage}
\newacronym{VNA}{VNA}{vector network analyzer}
\newacronym{VSG}{VSG}{vector signal generator}
\newacronym{V2I}{V2I}{vehicular to infrastructure}
\newacronym{V2V}{V2V}{vehicle to vehicle communications}
\newacronym{V2X}{V2X}{vehicle to everything communication}
\newacronym{V2LC}{V2LC}{Vehicular visible light communication}
\newacronym{4-PPM}{4-PPM}{4-Pulse Position Modulation}
\newacronym{VPPM}{VPPM}{variable pulse position modulation}
\newacronym{WSN}{WSN}{wireless sensor network}
\newacronym{QoS}{QoS}{quality of service}
\newacronym{SC-FDE}{SC-FDE}{Single Carrier Frequency Domain Equalization}
\newacronym{AWGN}{AWGN}{Additive White Gaussian Noise}
\title{Empirical Feasibility Analysis for Energy Harvesting Intra-Vehicular Wireless Sensor Networks}

\author{\IEEEauthorblockN{Mertkan Koca$^{1,2}$, Gokhan Gurbilek$^{1,2}$, Burak Soner$^{1,2}$, Sinem Coleri$^{1}$}
\IEEEauthorblockA{$^{1}$Department of Electrical and Electronics Engineering, Koc University, Sariyer, Istanbul, Turkey, 34450\\
$^{2}$Koc University Ford Otosan Automotive Technologies Laboratory (KUFOTAL), Sariyer, Istanbul, Turkey, 34450\\
E-mail: [mkoca14, ggurbilek13, bsoner16, scoleri]@ku.edu.tr}}

\begin{document}

\maketitle

\begin{abstract}

Vehicle electronic systems currently utilize wired networks for power delivery (from the main battery) and communication (e.g., LIN, CAN, FlexRay) between nodes. Wired networks cannot practically accommodate nodes in moving parts (e.g., tires) and with the increasing functional complexity in vehicles, they require kilometer-long harnesses, significantly increasing fuel consumption and manufacturing and design costs. As an alternative, energy harvesting \ac{IVWSN} can accommodate nodes in all locations and they obviate the need for wiring, significantly lowering costs. In this paper, we empirically analyze the feasibility of such an \ac{IVWSN} framework via extensive in-vehicle measurements for communications at 2.4 GHz, \ac{UWB} and \ac{mmWave} frequencies together with radio frequency (RF), thermal and vibration energy harvesting. Our analyses show that \ac{mmWave} performs best for short \ac{LoS} links in the engine compartment with performance close to \ac{UWB} for \ac{LoS} links in the chassis and passenger compartments in terms of worst case signal-to-interference-and-noise-ratio. For non-\ac{LoS} links, which appear especially more in the engine compartment and chassis, \ac{UWB} provides the highest security and reliability. 2.4 GHz suffers heavily from interference in all compartments while \ac{UWB} utilizes narrowband suppression techniques at the cost of lower bandwidth; \ac{mmWave} inherently experiences very low interference due to its propagation characteristics. On the other hand, \ac{RF} energy harvesting provides up to 1 mW of power in all compartments. Vibration and thermal energy harvesters can supply all nodes consuming $<$10 mW in the engine compartment and all $<$5 mW nodes in the chassis. In the passenger compartment, thermal harvesting is not available due to low temperature gradients but vibration and \ac{RF} sources can supply $<$1 mW nodes.

\end{abstract}

\begin{IEEEkeywords}
Energy harvesting, Intra-vehicular communication, Wireless Sensor Networks
\end{IEEEkeywords}

\IEEEpeerreviewmaketitle

\section{Introduction}

Electronic control systems have replaced their mechanical counterparts in modern vehicles since they can accommodate different functions with smaller design and integration cost \cite{navet_trends_in_auto_comm}. These systems contain networks of \ac{ECU} and many distributed sensors and actuators with wired interconnections for power (i.e., from the vehicle battery) and communication (e.g., LIN, CAN, FlexRay \cite{inVehComm_lin_can_flxry}). With the increasing demand for more functionality on vehicles, harnesses for these systems have reached up to 4 km of wiring and are still growing; this increases fuel consumption and manufacturing and design costs. Furthermore, these wired networks cannot practically accommodate nodes in moving parts (e.g. tire pressure monitoring and Intelligent Tire \cite{intel_tire}). A solution to these problems is converting these sensors and actuators into energy harvesting nodes which wirelessly communicate with each other and \ac{ECU}s \cite{scoleri_ivwsn_eh}.

Two main barriers towards realizing these energy harvesting intra-vehicular wireless sensor network (\ac{IVWSN}) are harvesting enough energy for their functionality \cite{eh_singleToMultiSrc}, and realizing reliable and secure communication within small delay at sufficient rate. The amount of harvestable energy and communication performance vary over both the respective technologies and the locations of nodes inside the vehicle. Therefore, to assess the feasibility of the \ac{IVWSN} framework, the potential of different energy harvesting and communication technologies need to be analyzed for node locations in different vehicle compartments (e.g., knock sensor in the engine, rain sensor in the body).

Existing studies on communication technologies for \ac{IVWSN} have focused on RFID (915 MHz), 2.4 GHz, ultra wideband (\ac{UWB}) and millimeter wave (\ac{mmWave}). Experimental studies on RFID \cite{rfid_ivwsn_ref} and 2.4 GHz \cite{4407229} have demonstrated the feasibility of communication in different vehicle compartments but power loss in passive RFID nodes and external interference into the vehicle at 2.4 GHz nodes are prohibitively large for safety-critical communications. Channel modeling studies for \ac{UWB} have have demonstrated the high reliability and energy efficiency of \ac{UWB} communication in all vehicle compartments. \cite{6679230,6289385,7447822}. Comparative studies between \ac{mmWave} and \ac{UWB} have shown that \ac{mmWave} provides better reliability for short line-of-sight links in the multipath-rich intra-vehicular environment due to higher attenuation of multipath components at higher frequencies \cite{7542517,5505745}. While these works extensively characterize communication technologies for the \ac{IVWSN} framework, in order to assess overall feasibility, a holistic evaluation which provides fair comparison of all respective communication and energy harvesting technologies is necessary. 

\begin{figure*}[!b]
\setlength\belowcaptionskip{0pt}
\centering
\includegraphics[width=17cm]{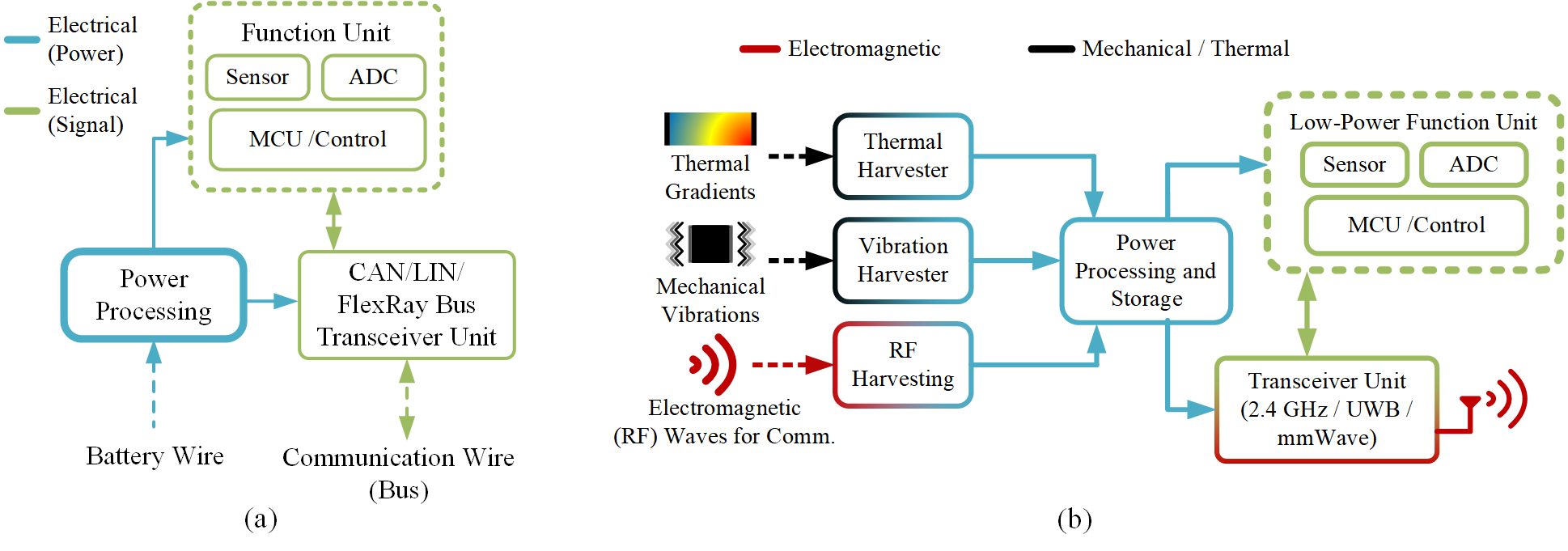}
\caption{Conventional vehicle sensor node (a) vs. energy harvesting \ac{IVWSN} node (b) architecture}
\label{wsn node architecture}
\end{figure*}

\begin{table*}[!b]
\setlength\belowcaptionskip{-0pt}
\centering
\begin{tabular}{|l|p{0.17\linewidth}|p{0.25\linewidth}|c|c|c|}
\hline
\textbf{Functional Domain}     & \textbf{Tasks} & \textbf{Examples and (Locations)} & \textbf{\begin{tabular}[c]{@{}c@{}}Power\\ {[}mW{]}\end{tabular}} & \textbf{\begin{tabular}[c]{@{}c@{}}Rate\\ {[}kbps{]}\end{tabular}} & \textbf{\begin{tabular}[c]{@{}c@{}}Security and\\ Reliability\end{tabular}} \\ \hline
Engine & Torque generation  & Knock (E), mass air flow (E), shaft torque (C) & $\sim$100 & 10 - 1000 & High \\ \hline
Powertrain & Transferring engine torque to the wheels  & Shaft torque (C), transmission speed (C) & $\sim$100 & 10 - 1000 & High \\ \hline
Chassis & Vehicle dynamics & Wheel speed (C), roll-stability (C)  &  $\sim$100 & 10 - 100 & High \\ \hline
Occupant safety  & Crash precautions & Seat belt (P), seat occupancy (P)  & 1 - 100 & 10 - 100 & High \\ \hline
Body     & General body control & Washer fluid (E), parking aid (C), windshield rain and fog (P) & 1 - 100 & \textless{}10 & Low \\ \hline
\end{tabular}
\caption{Requirements of vehicular sensors by functional domains. E: Engine, C: Chassis, P: Passenger compartments}
\label{t_node_req}
\end{table*}

The goal of this paper is to provide an empirical feasibility analysis for energy harvesting \ac{IVWSN} by evaluating the potential of \ac{RF}, vibration and thermal energy harvesting together with 2.4 GHz, \ac{UWB} and \ac{mmWave} communication technologies in chassis, engine and passenger compartments. A previous study \cite{scoleri_ivwsn_eh} surveys relevant data from prior art for the same technologies and compartments but lacks the empirical fair comparison of all technologies in all compartments, which is the main contribution of this paper. Section II presents the \ac{IVWSN} node model, and the power and communication requirements for vehicular sensor nodes. Section III compares the performances of 2.4 GHz, \ac{UWB} and \ac{mmWave} communication technologies via path loss and penetration loss measurements for 210, 182 and 156 links in the chassis, engine and passenger compartments, respectively. Section IV evaluates energy harvesting potential in the same vehicle compartments based on the path loss (for \ac{RF}), acceleration (for vibration) and temperature (for thermal) measurements from typical driving scenarios on a Fiat Linea MY 2009 and a Fiat Tipo 2016 using corresponding energy harvester models from literature. Section V provides an overall feasibility analysis of energy harvesting \ac{IVWSN}, considering the power consumption of vehicular sensors together with their energy harvesting and communication performance in different vehicle compartments and describes our related future work.

\section{Energy Harvesting \ac{IVWSN} Model \& Requirements} \label{requirement_section}

This section presents the energy harvesting \ac{IVWSN} node model and provides the associated power and wireless communication requirements.

\subsection{Node Model}

The conventional vehicular sensor node consists of three units: a power processing unit wired to the main battery, a wired vehicular network transceiver and a function unit which typically consists of \ac{MCU}, \ac{ADC} and sensing elements. The energy harvesting \ac{IVWSN} node is different from the conventional node in two aspects as demonstrated in Fig. \ref{wsn node architecture}: it utilizes energy harvesting to power itself rather than receiving power from the main battery and it employs a wireless communication transceiver rather than a wired one. Due to these differences, IVWSN nodes have power and communication requirements that are different from conventional nodes.

\subsection{Node Requirements}

There are five main functional domains in a vehicle: engine, powertrain, chassis, occupant safety and body. Power and communication requirements of conventional sensor nodes in these domains, which are distributed over the engine, chassis and passenger compartments on the vehicle, are shown in Table \ref{t_node_req}. \ac{IVWSN} node requirements are derived from these numbers with respect to two constraints: 1) power consumption due to limited harvested energy and 2) reliable and secure communication at a sufficient rate in the presence of noise and interference on the wireless channel. 

\subsubsection{Low Power Consumption}

To operate on the limited power harvested from available \ac{RF}, thermal and vibration sources in the vehicular environment, an \ac{IVWSN} node needs to be designed for very low power consumption; the feasible target is $<$10 mW \cite{sub10mwref}. There are three main approaches for lowering node power consumption: low-power sensor design, low-power transceiver design, and duty cycling. 

Low-power sensor design considers hardware-specific methods such as sub-threshold or low-voltage operation \cite{ultraLowPwrICDesign} and smart soft methods such as lowering the sensing rate when the measured signal is changing slowly \cite{nxp_whtPpr}. For transceivers, power consumption increases with higher rate and higher transmission power \cite{lowPowerRFDesign}. Therefore, low-power transceivers should employ the lowest possible rate and transmission power that satisfy communication requirements. Duty cycling can be employed for both sensors and transceivers; duty cycling refers to a node with a relaxed timing requirement switching itself into a low-consumption standby/sleep mode for the majority of its operational cycle  \cite{ultraLowPwr}. 

Example \ac{IVWSN} nodes from literature that utilize these three techniques have been demonstrated to consume $<$10 mW. A body inertial sensor that utilizes duty cycling and employs a low-power \ac{UWB} transceiver has been shown to consume only $\sim$5 mW \cite{zhang_2017}. Similarly, a wireless occupant safety sensor (seat occupancy) has been shown to consume $\sim$2 mW on average \cite{mds}. On the other hand, there aren't any current notable examples for engine, powertrain, chassis and occupant safety nodes with strict rate and timing requirements since they cannot effectively be duty cycled like the previous examples due to busy a operational cycle. However, those sensors can still benefit from special low-power sensor design and low-power transceiver design techniques to similarly lower their $\sim$100 mW power consumption to the target $<$10 mW level. 

\subsubsection{Wireless Communication} 

IVWSN transceivers are required to provide sufficient rate, reliability and security in spite of noise and interference on the wireless communication channel. The levels of rate, reliability and security required are determined by the sensing rate and the safety-criticalness of the sensed data.

Rate requirements are derived from bit resolution and sensing rate requirements of nodes. The highest rate requirements are for powertrain and engine nodes since they are subject to the fastest dynamics on the vehicle. Typically, they measure physical quantities at $<$20 bit resolution and $<$10kHz sensing rates, resulting in a raw rate requirement of at most 200 kbps. Since the nodes also share the wireless communication medium, the actual rate requirements for the engine and powertrain domains are higher, on the order of 1 Mbps. Moreover, chassis and occupant safety sensors have lower rate requirements (i.e., up to 100 kbps) since they are subject to the slower vehicle and the passenger dynamics. Finally, body sensors have the lowest rate requirements (i.e., up to 10 kbps) since they typically require very low bit resolution or simply have binary detection tasks.

Security and reliability pertains to attaining a low packet loss ratio under noise and interference on the wireless communication channel. The security and reliability requirement for an \ac{IVWSN} node is determined with respect to the severeness of packet loss for that node. Body sensors have a low security and reliability requirement since their tasks are not safety-critical (e.g., windshield fog detection, theft detection). On the other hand, safety-critical nodes in the engine, powertrain, chassis and occupant safety domains (e.g., knock, wheel speed, airbag) have a high security and reliability requirement since packet losses for those nodes can lead to fatal accidents.

\section{RF Communication for \ac{IVWSN}} \label{rfcom}

In this section, we evaluate the performance of 2.4 GHz, \ac{UWB} and \ac{mmWave} in terms of communication rate, reliability and security requirements of \ac{IVWSN} nodes by considering path and penetration loss measurements for node locations in the chassis, engine and passenger compartments of a Fiat Linea MY 2009 together with models of state-of-the-art transceivers from literature. The transceiver models consider the maximum achievable reliable communication rate, maximum allowed transmit power, transceiver noise level and power consumption metrics.

\begin{figure}[!b]
\setlength\belowcaptionskip{-10pt}
\centering
\includegraphics[width=9cm]{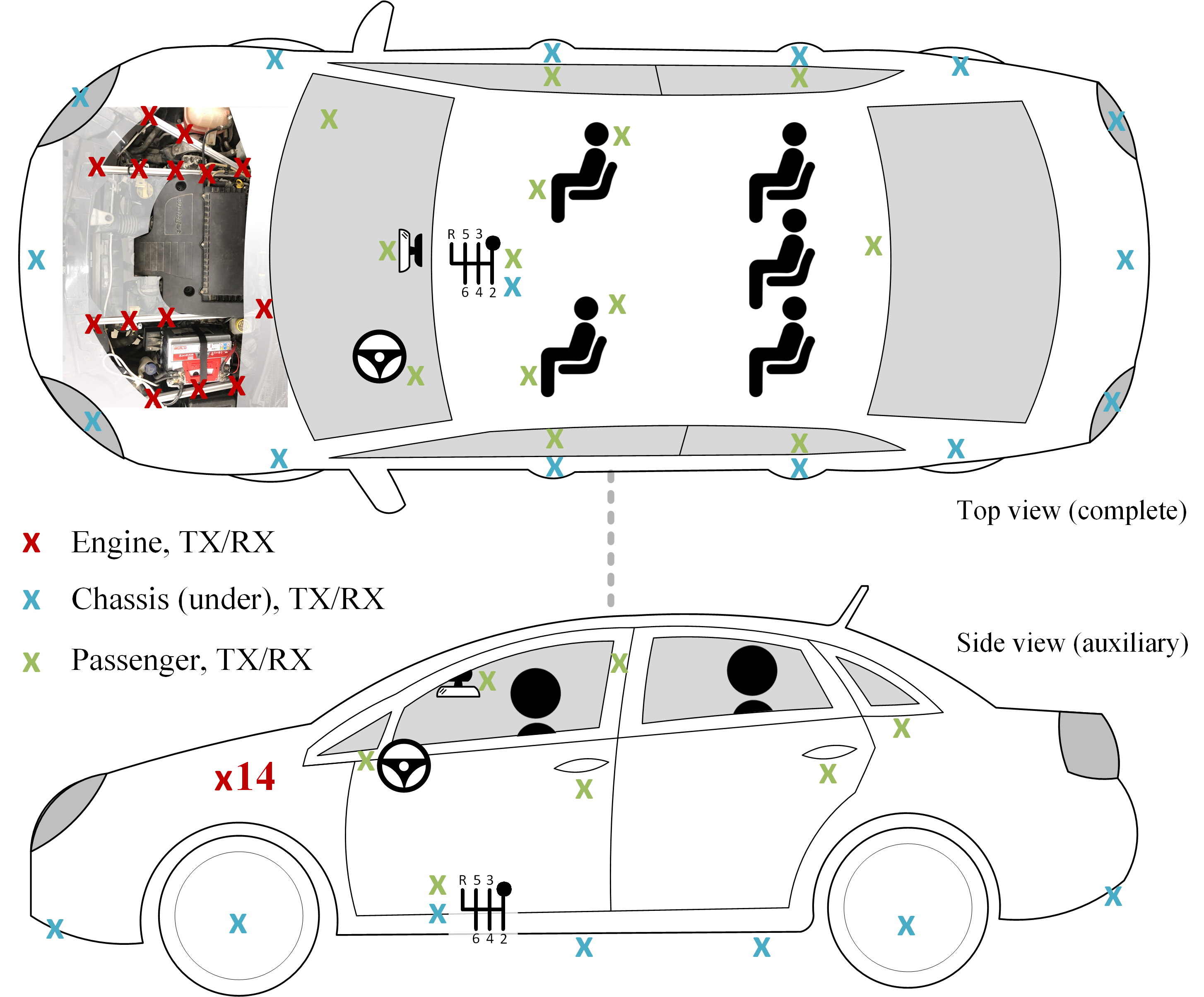}
\caption{Measurement locations on the Fiat Linea MY 2009}
\label{linea}
\end{figure}

\subsection{Rate}

Rate performance for each technology is evaluated by comparing the maximum achievable rates of the transceivers with the rate requirements of \ac{IVWSN} nodes for each domain (provided in Table \ref{t_node_req}). While state-of-the-art transceivers for 2.4 GHz have been shown to provide 0.25-54 Mbps \cite{24trans1,24trans2,24trans3}, \ac{mmWave} \cite{mmwavetrans1,mmwavetrans2,mmwavetrans3} and \ac{UWB} \cite{uwbtrans1,uwbtrans3} transceivers generally provide rates higher than 3000 Mbps, up to a maximum of $\sim$7000 Mbps. Therefore, all technologies have transceiver examples that are capable of at least an order of magnitude higher rates than the associated requirements in Table \ref{t_node_req}. For our later analyses, we choose one of these examples for each technology as our transceiver model. Regarding the low power consumption requirement and the rate versus power consumption trade-off for transceivers \cite{lowPowerRFDesign}, the transceiver that consumes the least power is chosen among options that satisfy the rate requirements. These are, \cite{24trans1} for 2.4 GHz, \cite{mmwavetrans1} for \ac{mmWave} and \cite{uwbtrans3} for \ac{UWB}.

\begin{figure}[!b]
\setlength\belowcaptionskip{-0pt}
\centering

    \begin{minipage}[b]{0.495\textwidth}
    \centering
    \includegraphics[width=\textwidth]{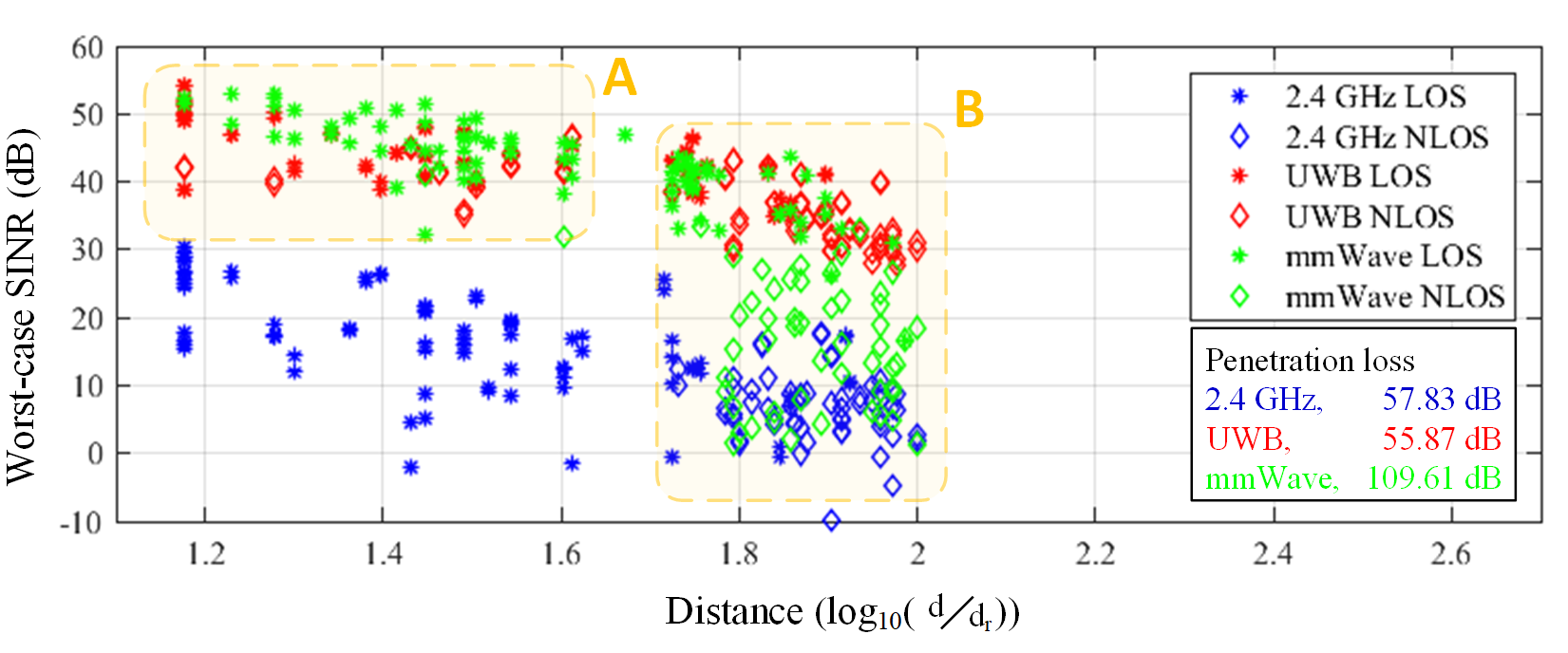}
    \caption*{(a)}
    \end{minipage}%

    \hspace{3mm}
    \begin{minipage}[b]{0.495\textwidth}
    \centering
    \includegraphics[width=\textwidth]{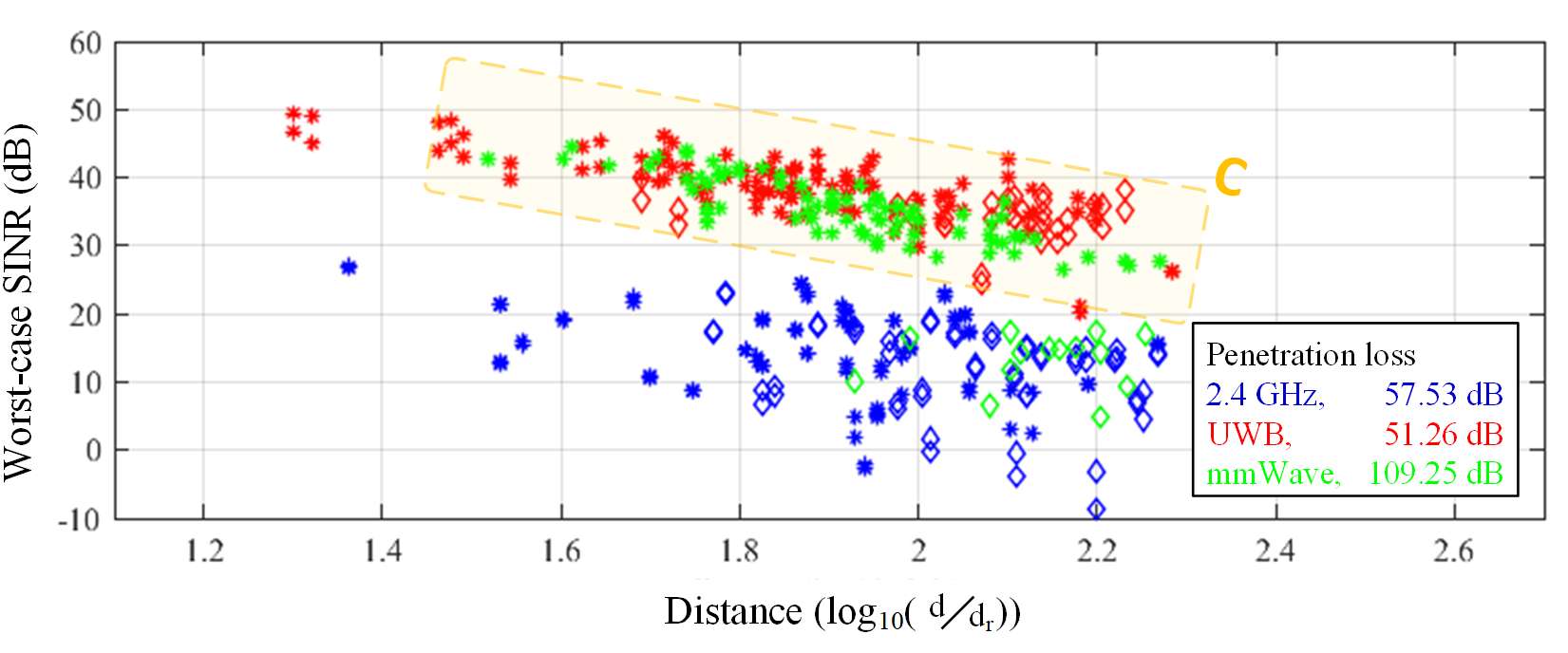}
    \caption*{(b)}
    \end{minipage}

    \hspace{3mm}
    \begin{minipage}[b]{0.495\textwidth}
    \centering
    \includegraphics[width=\textwidth]{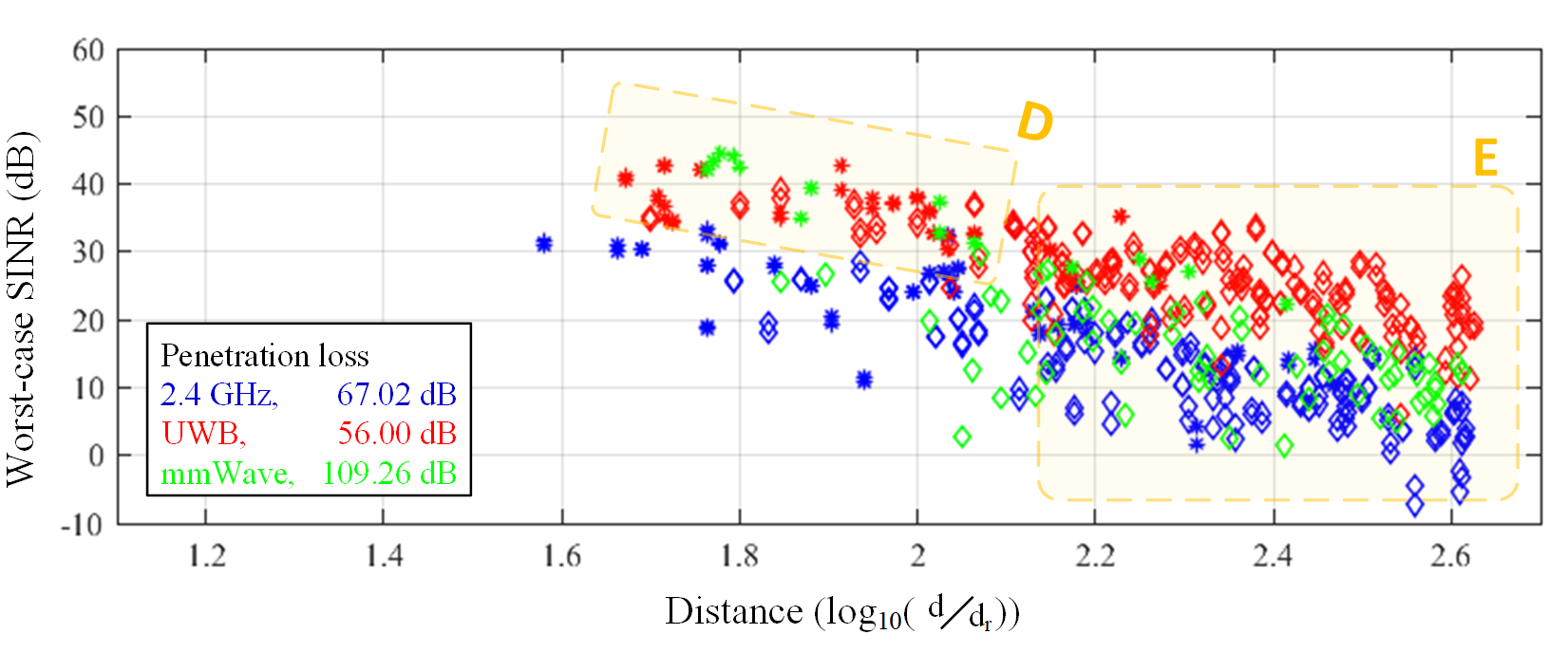}
    \caption*{(c)}
    \end{minipage}    

\caption{Worst-case \ac{SINR} for links in the (a) Engine compartment (b) Passenger compartment (c) the chassis}
\label{rf_sinr}
\end{figure}

\subsection{Security and Reliability}

Security and reliability performance analysis is based on the communication performance in the presence of noise and interference. For fair comparison, since the packet loss is determined by the \ac{SINR}, we evaluate the worst-case \ac{SINR} for each link, for each technology, in all compartments. The worst-case \ac{SINR} for a link occurs with maximum interference power which considers the maximum interferer \ac{TX} power and the minimum penetration loss for that compartment and technology. While path and penetration loss are measured, maximum \ac{TX} powers for both the interferer and the communicator, and the noise power are determined from transceiver specifications and associated standards.

Penetration loss was measured by placing a receiver (RX) antenna in each compartment and recording received power levels for transmitter (TX) antenna locations on a $\sim$1m perimeter around the RX antenna. Path loss measurements were conducted for 182, 156 and 210 different links for the chassis, engine and passenger compartments, respectively (Fig. \ref{linea}). Using the path and penetration loss measurements, the maximum transmit power levels (52, 60 and -11.3 dBm including 30, 50 and 0 dBi antenna gains for 2.4 GHz, \ac{mmWave} and \ac{UWB}, respectively \cite{scoleri_ivwsn_eh,ecfr2020}) and the selected low-power transceiver noise levels (-76 dBm for 2.4 GHz \cite{24trans1}, -68 dBm for \ac{mmWave} \cite{mmwavetrans1} and -84 dBm for \ac{UWB} \cite{uwbtrans3}) we compute the worst-case \ac{SINR} for each link, for each technology, in all compartments, as shown in Fig. \ref{rf_sinr}, where the ”reference distance”, dr, is 1 cm. The minimum penetration loss values used in the worst-case \ac{SINR} computations for each technology in each compartment are also provided in Fig. \ref{rf_sinr}.

The results in Fig. \ref{rf_sinr} show that \ac{mmWave} performs best for the short \ac{LoS} links in the engine compartment (sector A) due to high directionality and low interference, but \ac{UWB} surpasses \ac{mmWave} for the non-\ac{LoS} links due to lower attenuation (sector B). For \ac{LoS} links in the passenger compartment (sector C) and the chassis (sector D), which are inherently longer distance, \ac{mmWave} faces higher attenuation compared to \ac{UWB} and thus the two perform similarly. However, majority of the links in the chassis are non-\ac{LoS}; \ac{UWB} therefore performs best in the chassis (sector E). Both \ac{UWB} and 2.4 GHz receive high interference from external sources in all compartments due to low penetration losses. While 2.4 GHz suffers heavily from this, \ac{UWB} utilizes narrowband interference suppression techniques at the cost of having a lower bandwidth, converging its \ac{SINR} towards \ac{SNR} (e.g., narrow-band interference suppression with adaptive notch filters \cite{arslan2006ultra}). \ac{mmWave} inherently experiences very low inteference since \ac{mmWave} interferers face very high penetration losses in all compartments.

\subsection{Feasible \ac{RF} Communication for \ac{IVWSN}}

Overall, regarding the above results and the communication requirements of IVWSN nodes provided in Table I, the rate requirement for all nodes in all compartments can be satisfied by all technologies. Evaluations on the worst-case \ac{SINR} levels for all links, for technologies, in all compartments show that \ac{mmWave} provides the highest security and reliability for short \ac{LoS} links in the engine compartment and is comparable to \ac{UWB} for \ac{LoS} links in the other compartments. For non-\ac{LoS} links, \ac{UWB} is the best performer in all compartments. While \ac{UWB} and 2.4 GHz both experience high interference due to low penetration loss, by sacrificing some of its bandwidth \ac{UWB} employs narrowband interference suppression techniques and preserves its high \ac{SINR}; 2.4 GHz suffers heavily and therefore is not secure. \ac{mmWave} inherently experiences very low interference due to very high penetration losses in its frequency band.

\section{Energy Harvesting for \ac{IVWSN}}

In this section, we evaluate the potential for energy harvesting in the engine, chassis and passenger compartments. To this end, we apply recorded temperature and acceleration data from a Fiat Tipo 2016 during typical driving scenarios and path loss measurements from the previous section to harvester models from the literature. The energy harvester models consider both the harvester dynamics and power processing efficiencies to estimate their useful power outputs. 

\begin{figure}[!t]
\setlength\belowcaptionskip{-10pt}
\centering

    \begin{minipage}[b]{0.5\textwidth}
    \centering
    \includegraphics[width=\textwidth]{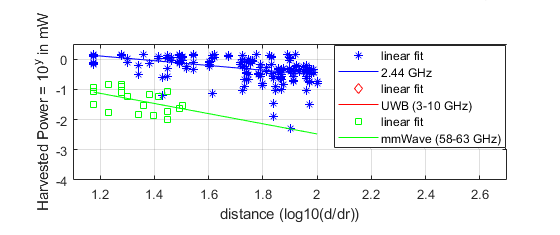}
    \caption*{(a)}
    \end{minipage}%

    \hspace{3mm}
    \begin{minipage}[b]{0.5\textwidth}
    \centering
    \includegraphics[width=\textwidth]{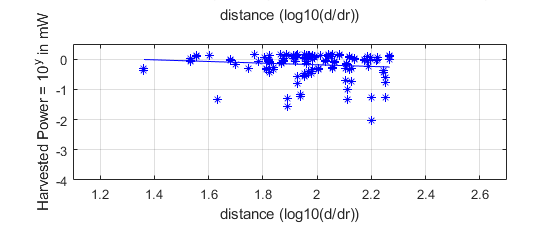}
    \caption*{(b)}
    \end{minipage}

    \hspace{3mm}
    \begin{minipage}[b]{0.5\textwidth}
    \centering
    \includegraphics[width=\textwidth]{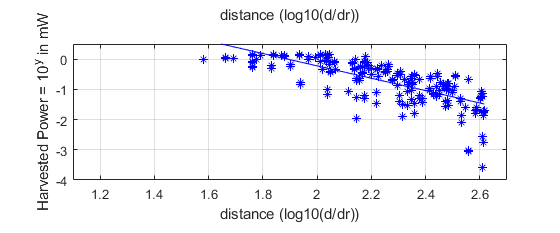}
    \caption*{(c)}
    \end{minipage}    

\caption{Amount of energy harvested (a) in Engine compartment (b) in Passenger compartment (c) in the Chassis}
\label{rf_eh}
\end{figure}

\subsection{\ac{RF} Sources}

\ac{RFEH} is used to convert part of the \ac{EM} energy in received \ac{RF} communication signals to useful power. Consisting of antennas, rectifiers, matching and switching circuits, the useful power output of an \ac{RFEH} system depends on the received power and the component efficiencies. Reported efficiencies for state-of-the-art harvesters are 11.5\% - 40\% for [-10, 18]-dBm input power at a sensitivity of -20 dBm for 2.4 GHz \cite{244,effs}, 5\% - 10\% for [-36, -25]-dBm input power at a sensitivity of -36 dBm for \ac{UWB} \cite{UWB} and peak 12\% with a sensitivity of 2 dBm for \ac{mmWave} \cite{mmWave,effs}. 

We estimate the actual harvestable power outputs in each compartment for each technology by applying the received power measurements from the previous section and the above mentioned efficiency levels from literature to our simulated \ac{RFEH} model. The results demonstrate two feasible options as shown in Fig. \ref{rf_eh}: up to $\sim$1 mW can be harvested from 2.4 GHz in all compartments, and up to $\sim$0.1 mW can be harvested from \ac{mmWave} in only the engine compartment.

\subsection{Vibration Sources}

Vibration energy harvester generates useful electrical power from the ambient vibrations. In an \ac{IVWSN} framework, this task can best be achieved with magnet-coil type nonlinear \ac{EMH} since they provide broad-band harvesting performance, and can thus accommodate for the typical varying frequency and high impact ($\sim$1-10 g on parts weighing kilograms) vibrations in a vehicle. Other choices such as piezoelectric and electrostatic harvesters are not suitable due to their low durability and design for smaller vibrations \cite{vibr_eh_rev_1}. 

In order to estimate the potential power harvestable by such an \ac{EMH}, we recorded acceleration data (with an ADXL345 accelerometer) during typical driving scenarios (with a Fiat Tipo 2016) from the engine block, the rear pillar inside the passenger compartment and the transmission unit under the chassis. We applied the measured acceleration data to a state-of-the-art \ac{EMH} model \cite{maglev_vibr_eh} in simulation environment and estimated the raw power output. The model considers the transfer characteristics of the harvester, in which raw output power is proportional to the square of the input acceleration. However, the raw output of the \ac{EMH} needs to be processed by additional power electronics to generate useful power; this typically adds a $\sim$50\% efficiency factor to the power output \cite{boost_conv,mitcheson2007power}. The useful electrical power output of the \ac{EMH} is shown Fig. \ref{sensorharv} (a): mean 170 \textmugreek W and 4.12 mW peak for the passenger compartment, mean 2.58 mW and peak 71.2 mW for the engine compartment and mean 1.13 mW peak 7.62 mW for the chassis.

\subsection{Thermal Sources}

\begin{figure}[t]
\setlength\belowcaptionskip{-10pt}
\centering

    \begin{minipage}[b]{0.5\textwidth}
    \centering
    \includegraphics[width=8.5cm]{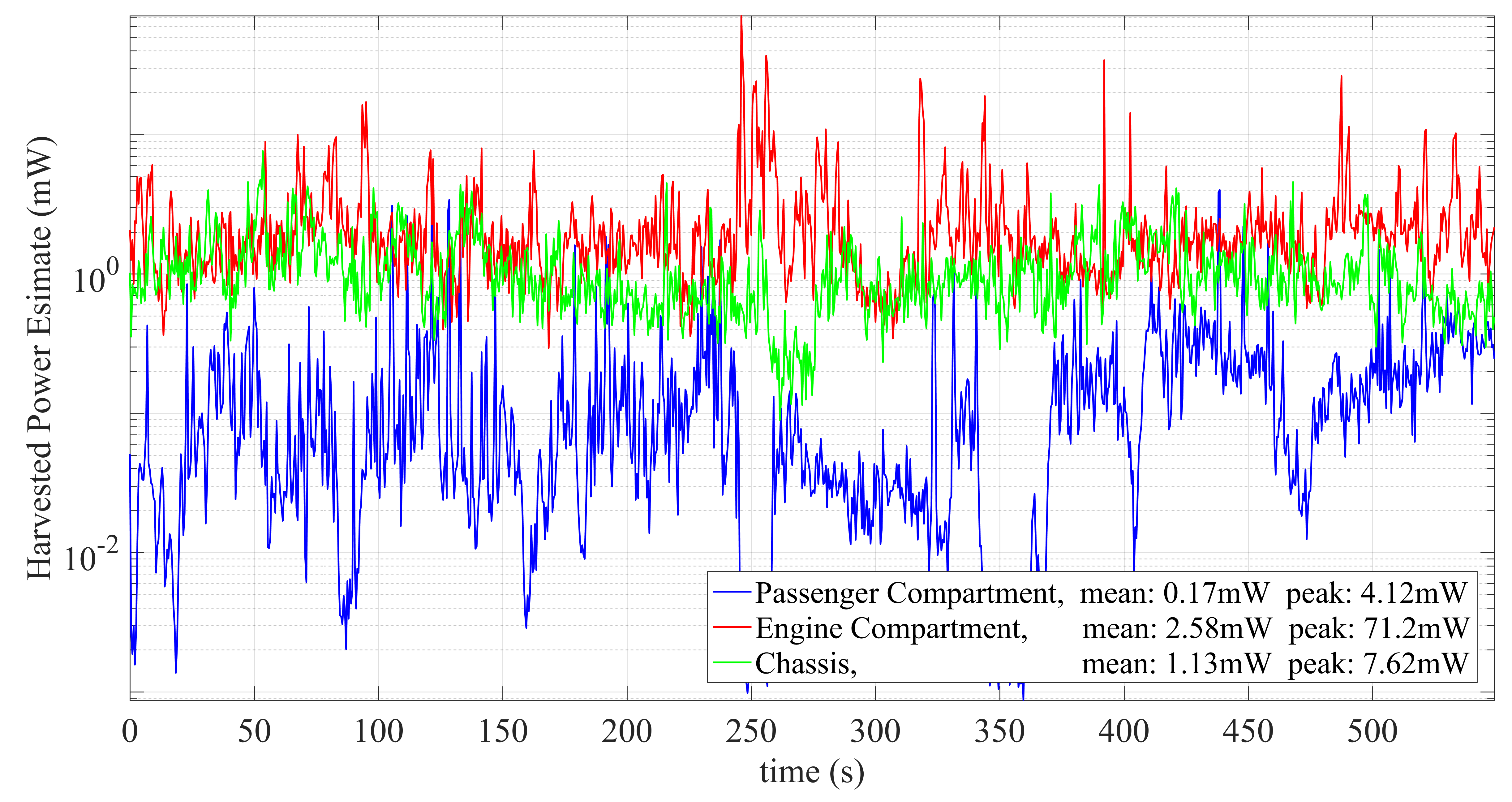}
    \caption*{(a)}
    \end{minipage}%

    \begin{minipage}[b]{0.5\textwidth}
    \centering
    \includegraphics[width=8.5cm]{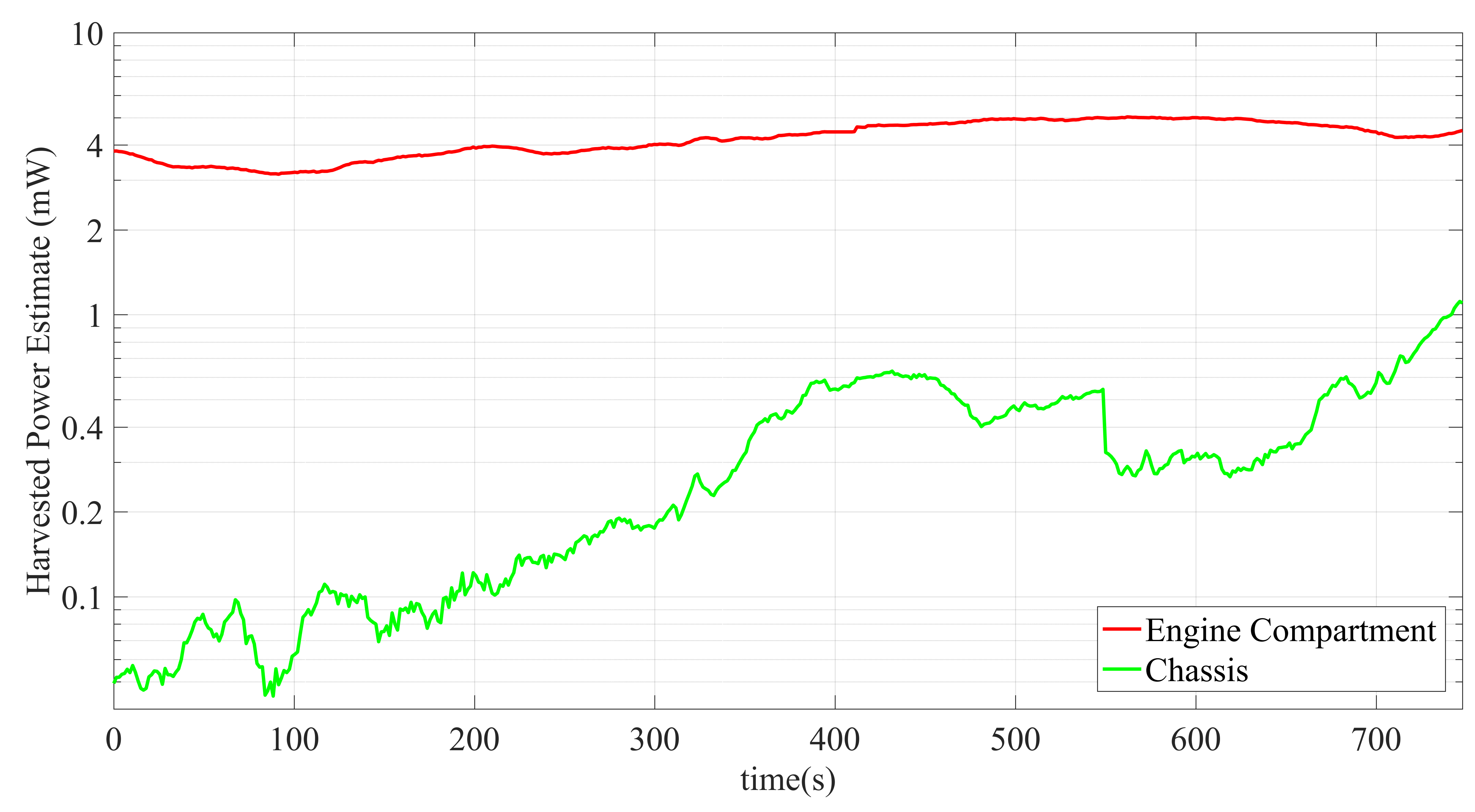}
    \caption*{(b)}
    \end{minipage}%

\caption{Vibration energy harvesting results (a) and thermal energy harvesting results (b)}
\label{sensorharv}
\end{figure}

\ac{TEG} convert the heat flow due to temperature difference between their two sides to DC electrical power. \ac{TEG} power output can be estimated from temperature measurements of these two sides. We collected simultaneous hot surface and ambient temperature data with K-type thermocouples and a MAX31856 amplifier module from the hot exit of the coolant pipe from the engine block, the rear pillar inside the passenger compartment and the transmission unit under the chassis on the same vehicle, during the same driving conditions as the vibration case. We applied the measured temperature data to a \ac{TEG} model \cite{thm_building} and estimated the raw power output. The model considers the transfer characteristics of the harvester, in which raw output power is proportional to the square of the temperature difference on its two sides. For generating useful power, the low voltage DC raw \ac{TEG} output also needs additional power processing circuitry, which brings a $\sim$30\% efficiency factor for the considered \ac{TEG} model \cite{thm_building}. The passenger compartment measurements showed temperature differences around 2\degree C, which is below the minimum $\sim$10\degree C difference required for harvesting \cite{thm_building,enescu2019thermoelectric}. Therefore, thermal energy harvesting inside the passenger compartment is not feasible. The useful electrical power output of the \ac{TEG} is shown in Fig. \ref{sensorharv} (b): $\sim$4 mW for the engine compartment (after the engine heats up) and on average 0.5 mW for the chassis.

\subsection{Feasible Energy Harvesting for \ac{IVWSN}}

Overall, regarding the above results and the requirements for low-power \ac{IVWSN} nodes in Section II, vibration, thermal and \ac{RF} energy harvesters can generate enough power to supply all nodes consuming $<$10 mW in the engine compartment and all $<$5 mW nodes in the chassis. While \ac{RF} and thermal sources provide stable outputs, vibration harvesters can occasionally peak at very high levels (e.g., engine, 71.2 mW) which can be stored for later use. In the passenger compartment, thermal sources cannot provide significant power due to low temperature gradients and vibration sources can support \ac{RF} sources with $<$1 mW on average due to low amplitude vibrations. 

\section{Conclusion}

\ac{IVWSN}s enable vehicular sensor nodes inside inaccessible locations such as tires and they free the vehicle of the wiring harness, significantly lowering cost and fuel consumption. We provide an empirical feasibility analysis of such an \ac{IVWSN} framework by evaluating the potential of 2.4 GHz, \ac{UWB} and \ac{mmWave} communication and \ac{RF}, vibration and thermal energy harvesting technologies for the chassis, engine and passenger compartments 

Analysis of state-of-the-art transceivers for 2.4 GHz, \ac{UWB} and \ac{mmWave} show that rate requirements for all \ac{IVWSN} nodes can be met by all technologies. Path and penetration loss measurements demonstrate that \ac{mmWave} provides the highest \ac{SINR} and thus the highest security and reliability for the short \ac{LoS} links in the engine compartment. However, due to higher attenuation over longer link distances, \ac{mmWave} is comparable to \ac{UWB} for \ac{LoS} links in the other two compartments. For non-\ac{LoS} links in all compartments and especially in the chassis, \ac{UWB} provides the highest \ac{SINR} due to lower attenuation. 2.4 GHz is not secure since it suffers heavily from interference. While \ac{UWB} is also vulnerable, it utilizes narrowband interference suppression techniques in exchange for lower bandwidth. \ac{mmWave} inherently experiences very low interference due to very high penetration loss in all compartments. 

For \ac{RF} energy harvesting, 2.4 GHz is the best performer, providing 0.1-1 mW useful power for most links in all compartments; \ac{UWB} cannot provide any useful power and \ac{mmWave} can only provide up to 0.1 mW only in the engine compartment. Vibration and thermal harvesting can support \ac{RFEH} to power all \ac{IVWSN} nodes that have $<$10 mW consumption in the engine compartment and all nodes that have $<$5 mW consumption in the chassis. Thermal sources cannot provide significant power in the passenger compartment due to insufficient temperature gradients, and vibration sources can support \ac{RF} sources with only $<$1 mW on average due to low vibration levels. While these analyses empirically confirm the feasibility of the \ac{IVWSN} framework from power and communication perspectives in different vehicle compartments, as future work, we aim to conduct an experimental study with actual low-power \ac{IVWSN} nodes with harvesters and low-power transceivers and sensors.

\bibliographystyle{ieeetr}
\bibliography{references}
\nocite{*}

\end{document}